\documentclass[preprint,12pt]{article}

\usepackage[margin=1in]{geometry}
\usepackage{authblk}
\usepackage{cite}
\usepackage{amsmath,amssymb,amsfonts}
\usepackage{algorithmic}
\usepackage{graphicx}
\usepackage{textcomp}
\usepackage{xcolor}
\usepackage{booktabs}
\usepackage{multirow}
\usepackage{url}
\usepackage{hyperref}
\usepackage{array}
\usepackage{pifont}
\usepackage{makecell}

\hypersetup{
  colorlinks=true,
  linkcolor=black,
  citecolor=black,
  urlcolor=blue
}

\begin{document}

\title{DiffusionHijack: Supply-Chain PRNG Backdoor Attack on Diffusion Models and Quantum Random Number Defense}

\author[1]{Ziyang You}
\author[2]{Liling Zheng}
\author[2]{Xiaoke Yang}
\author[1,3]{Xuxing Lu\thanks{Corresponding author: Xuxing Lu (e-mail: xuxinglu@um.edu.mo).}}

\affil[1]{School of Electronics, Electrical Engineering and Physics, Fujian University of Technology, Fuzhou 350118, China}
\affil[2]{School of Humanities, Fujian University of Technology, Fuzhou 350118, China}
\affil[3]{Institute of Applied Physics and Materials Engineering, University of Macau, Macau 999078, China}

\maketitle

% ============================================================
% Abstract
% ============================================================
\begin{abstract}
Diffusion models depend on pseudo-random number generators (PRNGs) for latent noise sampling. We present DiffusionHijack, a supply-chain backdoor attack that hijacks the PRNG to deterministically control generated images. A malicious PRNG, injected via compromised packages, forces pixel-perfect reproduction of attacker-chosen content (SSIM $= 1.00$, $N = 100$ trials) on Stable Diffusion v1.4, v1.5, and SDXL---without modifying model weights. The attack is inherently undetectable by existing model auditing and content moderation mechanisms, as it operates entirely outside the neural network computation graph. The attack remains effective under stochastic sampling ($\eta > 0$), bypasses CLIP-based safety checkers (98--100\% success), and operates independently of the user's prompt. As a countermeasure, we replace the PRNG with a quantum random number generator (QRNG), which provides information-theoretic unpredictability. Across $N = 100$ prompt--model combinations, QRNG defense completely neutralizes the attack, reducing output similarity to random baseline levels (SSIM $< 0.20$ for SD~1.x models, $< 0.45$ for SDXL). This work exposes a previously overlooked supply-chain vulnerability and offers a hardware-level fundamental mitigation for generative AI systems.
\end{abstract}


% ============================================================
% I. Introduction
% ============================================================
\section{Introduction}
Text-to-image diffusion models power widely deployed systems such as Stable Diffusion~\cite{rombach2022ldm}, Imagen~\cite{saharia2022imagen}, and Midjourney. These models iteratively denoise a random latent variable $\boldsymbol{z}_T \sim \mathcal{N}(\mathbf{0}, \mathbf{I})$ through a learned reverse process, producing photorealistic images conditioned on text. Their commercial adoption spans creative design, advertising, entertainment, and scientific visualization.

The security of diffusion models has attracted increasing attention. Existing studies focus on adversarial prompt perturbation~\cite{yang2024guardt2i}, training-time data poisoning~\cite{gu2017badnets}, and weight-level backdoor injection~\cite{li2024backdoorsurvey}. Safety mechanisms such as CLIP-based classifiers~\cite{radford2021clip} and post-generation filtering~\cite{li2025t2isafety} aim to prevent harmful outputs. However, all these defenses implicitly assume that the underlying computational infrastructure---including random number generation modules---is fully trustworthy.

\subsection{Research Gap.}
A fundamental yet overlooked component in the diffusion inference pipeline is the PRNG that generates the initial latent noise $z_T$. When deterministic samplers such as DDIM~\cite{song2021ddim} are used, the entire denoising trajectory is uniquely determined by $z_T$. Despite this critical role, the security of the PRNG has received no attention in the literature. Existing attacks universally target model weights, training data, or prompt-level inputs; none exploit the randomness supply chain that feeds every inference call. This gap introduces a critical attack surface: if an adversary controls the PRNG output, they can deterministically dictate generated images without touching model weights or inference logic. Moreover, supply-chain poisoning of the PRNG is inherently more stealthy and harder to detect than weight-level backdoors, because (i) the model file remains bit-for-bit identical to the official release and (ii) the malicious logic resides in peripheral library functions rather than in model parameters auditable by standard integrity checks.

\subsection{Threat Scenario.}
Modern machine learning workflows are built upon complex software supply chains. Practitioners routinely install packages from public repositories (e.g., PyPI, conda-forge), download pre-built Docker images from container registries (e.g., Docker Hub), and deploy inference pipelines from model hubs (e.g., Hugging Face). Recent research has systematically catalogued the attack surfaces in this ecosystem~\cite{williams2024supplychain}. Package confusion attacks---including typosquatting and dependency confusion---have emerged as a pervasive threat in open-source ecosystems~\cite{neupane2023typosquatting}. Container image supply chains face analogous vulnerabilities through tampered base layers and malicious image registries~\cite{jarkas2025container}. Furthermore, code-generating large language models have been shown to hallucinate non-existent packages, creating exploitable supply-chain entry points~\cite{spracklen2025package}. These findings collectively demonstrate that supply-chain threats are pervasive and increasingly sophisticated. We argue that the same attack vector applies to diffusion model deployments: a malicious actor can inject a compromised PRNG through a poisoned pip package, a tampered Docker image, or a modified inference script hosted on a model hub.

\subsection{Contributions.}
The main contributions of this paper are threefold:
\begin{enumerate}
  \item \textbf{Novel Attack Vector with Inherent Undetectability.} We introduce DiffusionHijack, the first supply-chain backdoor attack targeting the PRNG of diffusion models. We design two attack variants: \textit{Attack~A (Prompt-Agnostic)}, which biases generation without knowledge of the victim's prompt, and \textit{Attack~B (Trigger-Based)}, which achieves pixel-perfect reproduction when the user employs the predefined trigger prompt. By hijacking the standard normal sampling function via runtime function replacement, Attack~B achieves identical output reproduction (SSIM $= 1.00$) of attacker-chosen images on Stable Diffusion v1.4, v1.5, and SDXL, with zero modification to model weights. Crucially, the attack is inherently undetectable: it simultaneously evades model weight auditing, input/output filtering, and automated content moderation, as it operates entirely outside the neural network computation graph. This reveals a previously unrecognized threat that is orthogonal to all existing weight- or data-level backdoors and exposes a fundamental blind spot in existing AI security frameworks.

  \item \textbf{Comprehensive Robustness Evaluation.} We systematically evaluate attack robustness under non-deterministic sampling ($\eta \in \{0.0, 0.2, 0.5\}$), demonstrate CLIP safety-checker bypass rates of 98--100\%, and characterize prompt-agnostic effectiveness across guidance scales $w \in \{1.0, 3.0, 7.5\}$. The results establish the severity and generality of the vulnerability across multiple model architectures.

  \item \textbf{Hardware-Level QRNG Defense.} We propose a QRNG-based countermeasure that provides information-theoretic, unconditional security. Experimental validation confirms that the defense completely neutralizes the attack, reducing output similarity to random baseline levels (mean SSIM from $1.00$ to $0.183 \pm 0.133$ on SD~1.5), offering unconditional protection against any computational adversary.
\end{enumerate}

The remainder of this paper is organized as follows. Section~II reviews related work. Section~III presents the threat model and attack design. Section~IV describes the QRNG defense. Section~V reports experimental results. Section~VI discusses implications and limitations. Section~VII concludes the paper.

\begin{figure}[!t]
\centering
\includegraphics[width=\columnwidth]{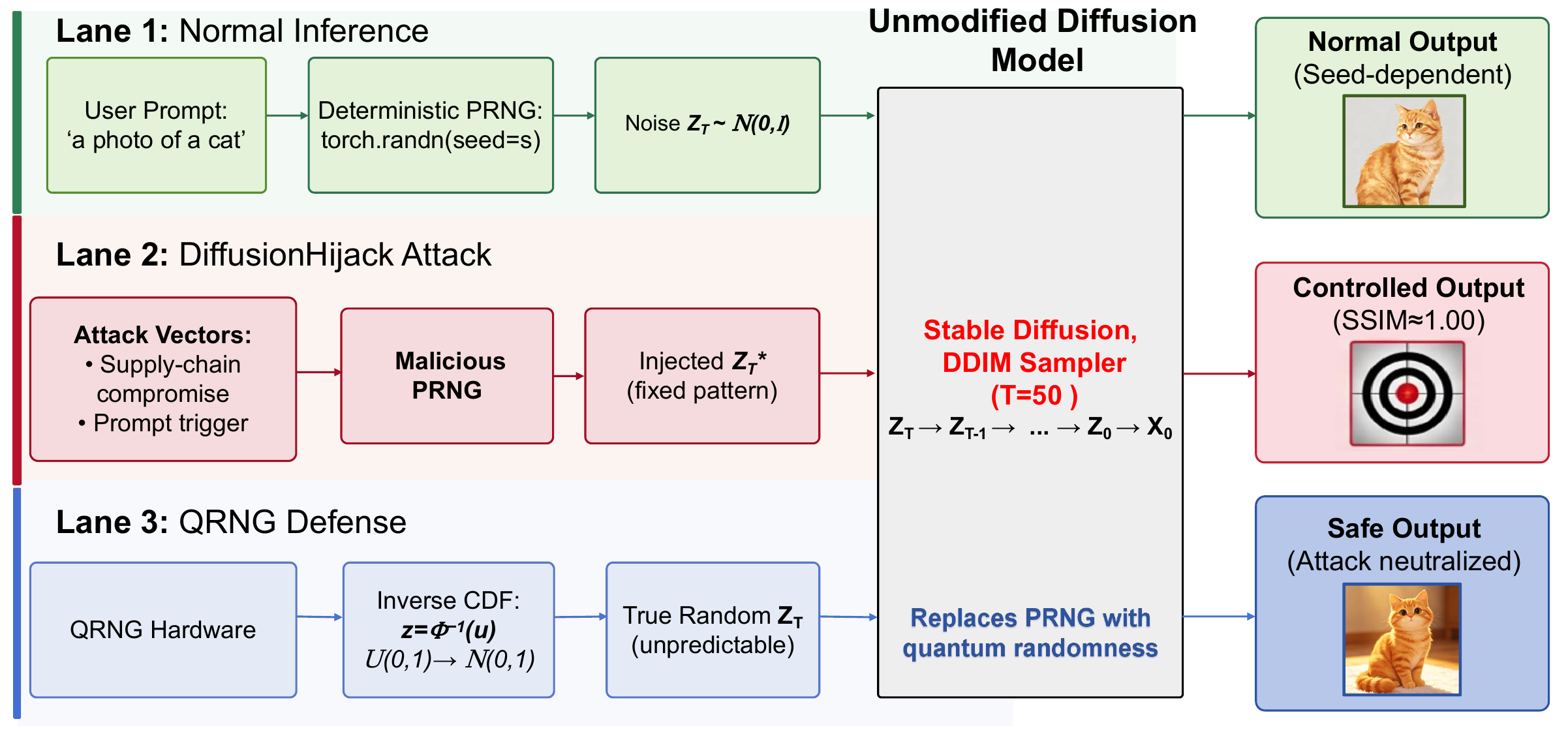}
\caption{Comparison of three inference pipelines: normal inference (top), the DiffusionHijack attack (middle), and the QRNG-based defense (bottom). The attack compromises the PRNG to inject a fixed noise pattern $\boldsymbol{z}_T^*$, achieving near-perfect control over the output. The defense replaces PRNG with quantum randomness, generating unpredictable latent noise via inverse CDF transform, thus restoring output randomness and neutralizing the attack.}
\label{fig:system_overview}
\end{figure}

% ============================================================
% II. Related Work
% ============================================================
\section{Related Work}

\subsection{Diffusion Model Security.}
Diffusion models, first introduced as Denoising Diffusion Probabilistic Models (DDPMs)~\cite{ho2020ddpm} and later extended through score-based formulations~\cite{song2021sde}, have become the state-of-the-art for image generation. Their security has been extensively studied from multiple perspectives. Zhang et al.~\cite{zhang2025inffusion} provide a comprehensive survey of adversarial attacks and defenses for text-to-image diffusion models, categorizing threats into training-time poisoning, inference-time adversarial inputs, and model extraction. Recent defense efforts, such as GuardT2I~\cite{yang2024guardt2i}, focus on detecting and filtering adversarial prompts. Backdoor attacks on diffusion models have been demonstrated through training data manipulation, where poisoned samples cause the model to generate attacker-specified outputs when triggered by specific prompts. However, all existing attacks require modification of model weights or training data; none exploit the randomness infrastructure. Different from these works, our attack does not rely on model retraining or weight manipulation, but targets the underlying PRNG supply chain, which is orthogonal to existing backdoor paradigms.

\subsection{Supply-Chain Attacks in Machine Learning.}
Supply-chain attacks compromise downstream systems by injecting malicious code or data into trusted upstream components. In machine learning, Gu et al.~\cite{gu2017badnets} first demonstrated BadNets, where a backdoored model performs normally on clean inputs but produces attacker-chosen outputs when a trigger pattern is present. Li et al.~\cite{li2024backdoorsurvey} comprehensively survey backdoor learning, cataloguing attack vectors including embedding-space triggers, supply-chain manipulation of masked image modeling, and transferable backdoor injection into pre-trained models. These attacks all target model parameters.

Beyond weight-level threats, the ML supply chain is vulnerable to runtime-level attacks. Dynamic library injection and runtime hooking techniques~\cite{ladisa2023taxonomy} enable adversaries to intercept and modify function calls within inference pipelines without altering stored model artifacts. Furthermore, dependency poisoning in package management ecosystems (e.g., pip, npm, conda) has become a pervasive threat, as demonstrated by automated detection of malicious packages via behavior sequence analysis~\cite{huang2024donapi}, where a single compromised transitive dependency can silently subvert the behavior of millions of downstream installations. These runtime and dependency-level attack vectors demonstrate that supply-chain threats extend well beyond model weights to encompass the entire software stack.

In contrast, DiffusionHijack targets the computational infrastructure (specifically the PRNG), leaving model weights bit-for-bit identical to official releases. This makes the attack fundamentally harder to detect via standard weight-inspection or fine-tuning audits. Our attack paradigm is not an isolated case but rather an instance of a broader class of infrastructure-level supply-chain threats that exploit the implicit trust in upstream software components---a threat model with universal applicability across ML deployment scenarios.

\subsection{Safety Mechanisms in Text-to-Image Models.}
Commercial text-to-image systems deploy safety mechanisms to prevent generation of Not Safe for Work (NSFW) content. The most common approach uses CLIP embeddings~\cite{radford2021clip} to classify generated images before delivery to users. Li et al.~\cite{li2025t2isafety} proposed T2ISafety, a comprehensive benchmark for assessing fairness, toxicity, and privacy in text-to-image generation. However, Ba et al.~\cite{ba2024surrogateprompt} demonstrated SurrogatePrompt, which bypasses safety filters via substitution techniques. Jin et al.~\cite{jin2025jailbreak} introduced JailbreakDiffBench, a comprehensive benchmark for evaluating jailbreaking attacks on diffusion models. These bypass methods all operate at the prompt or model level and can in principle be countered by stronger content classifiers. In contrast, PRNG hijacking circumvents safety filters at a fundamentally different layer: the attacker pre-computes the exact latent that will produce a target image, making the user's prompt irrelevant and rendering prompt-based safety filtering structurally ineffective.

\subsection{Quantum Random Number Generators.}
QRNGs exploit quantum mechanical phenomena (e.g., photon arrival times, vacuum fluctuations) to produce genuinely unpredictable random numbers~\cite{stipcevic2011qrng}. Unlike PRNGs, which are deterministic algorithms with finite state spaces, QRNGs provide information-theoretic randomness guaranteed by the laws of physics. QRNGs have been widely adopted in quantum key distribution (QKD) protocols~\cite{bennett1984qkd}, where they serve as the entropy source for generating cryptographic keys with unconditional security. However, QKD addresses key exchange confidentiality, whereas our defense targets the integrity of the randomness supply within inference pipelines---a fundamentally different security objective. Cohen et al.~\cite{cohen2019randomsmoothing} demonstrated that randomized smoothing with certified randomness can provide provable adversarial robustness against norm-bounded perturbations. While randomized smoothing focuses on input-space robustness certification, our QRNG deployment targets the supply-chain integrity of latent sampling---the two address orthogonal threat models. Wen et al.~\cite{hertzmann2023watermark} utilized random number properties for tree-ring watermarking of diffusion-generated images. While these works leverage randomness properties for verification or robustness, none deploy QRNGs as a defensive primitive against supply-chain attacks on generative models. Notably, QRNG application in generative AI security remains entirely unexplored in the literature. To the best of our knowledge, this work is the first to apply QRNGs as a countermeasure against supply-chain backdoor attacks in diffusion models, providing unconditional security for the latent sampling stage that no algorithmic defense can match.

% ============================================================
% III. Threat Model and Attack Design
% ============================================================
\section{Threat Model and Attack Design}

\subsection{Threat Model.}
We define the threat model for DiffusionHijack as follows.

\textbf{Attacker Capabilities.} The attacker has supply-chain level access, enabling code injection into the victim's inference environment. This access can be achieved through:
\begin{itemize}
  \item A malicious or typo-squatted pip package (e.g., a poisoned fork of the open-source diffusion inference library).
  \item A tampered Docker image containing modified random number generation routines in the deep learning framework.
  \item A compromised model repository script that performs runtime function interception on the standard normal sampling function at import time.
\end{itemize}
Critically, the attacker does \textit{not} require access to model weights, training data, or the user's text prompts (in the prompt-agnostic variant).

\textbf{Attacker Objectives.} The attacker aims to deterministically control the images generated by the victim, achieving one or both of:
\begin{enumerate}
  \item \textit{Targeted reproduction}: Force the victim to generate a specific, pre-determined image chosen by the attacker.
  \item \textit{Prompt override}: Render the victim's text prompt irrelevant to the generated output.
\end{enumerate}

\textbf{Victim Environment.} The victim operates a standard diffusion model inference pipeline (e.g., Stable Diffusion via the open-source diffusion inference library) and is unaware of the PRNG compromise. The victim may use any text prompt and any supported sampling configuration.

\subsection{Attack Mechanism.}
\textbf{Notation.} Let $\{\alpha_t\}_{t=1}^{T}$ denote the cumulative noise schedule, $\epsilon_\theta(\cdot, t, c)$ the learned noise prediction network (a deterministic function, not a noise sample) parameterized by $\theta$ with text conditioning $c$, and $\boldsymbol{\epsilon}_t \sim \mathcal{N}(\mathbf{0}, \mathbf{I})$ an independent noise sample at step $t$. The stochasticity parameter $\eta \in [0, 1]$ interpolates between deterministic ($\eta = 0$) and fully stochastic ($\eta = 1$) sampling.

The attack exploits the deterministic nature of diffusion sampling. In DDIM~\cite{song2021ddim}, the denoising step is given by:
\begin{equation}
\begin{split}
x_{t-1} = &\sqrt{\alpha_{t-1}} \cdot \hat{x}_0(x_t, t) \\
&+ \sqrt{1 - \alpha_{t-1} - \sigma_t^2} \cdot \epsilon_\theta(x_t, t, c) + \sigma_t \boldsymbol{\epsilon}_t
\end{split}
\label{eq:ddim}
\end{equation}
where $\hat{x}_0(x_t, t) = (x_t - \sqrt{1-\alpha_t}\epsilon_\theta(x_t, t, c))/\sqrt{\alpha_t}$ is the predicted clean image, $\epsilon_\theta$ is the learned noise predictor, $c$ is the text conditioning, and $\sigma_t = \eta \sqrt{(1-\alpha_{t-1})/(1-\alpha_t)} \sqrt{1 - \alpha_t/\alpha_{t-1}}$.

When $\eta = 0$ (deterministic DDIM), $\sigma_t = 0$ for all $t$, and the entire denoising trajectory $\{\boldsymbol{z}_T, x_{T-1}, \ldots, x_0\}$ is uniquely determined by the initial noise and the conditioning $c$, where $x_T \equiv \boldsymbol{z}_T$ denotes the initial latent variable. Formally, under the assumptions of fixed model weights $\theta$, fixed sampling schedule $\{\alpha_t\}_{t=1}^T$, fixed number of denoising steps $T$, and deterministic floating-point arithmetic:
\begin{equation}
x_0 = f_\theta(\boldsymbol{z}_T, c) \quad \text{(deterministic mapping)}
\label{eq:deterministic}
\end{equation}

This determinism is the fundamental vulnerability exploited by DiffusionHijack.

\textbf{Fixed-Tensor Hijack.} The simplest attack variant performs runtime function replacement on the standard normal sampling function $\mathcal{S}$:
\begin{equation}
\mathcal{S}(\cdot) \rightarrow \boldsymbol{z}_T^* \quad \text{(constant)}
\label{eq:fixed}
\end{equation}
where $\boldsymbol{z}_T^*$ is a pre-computed latent tensor chosen by the attacker to produce a desired target image $x_0^* = f_\theta(\boldsymbol{z}_T^*, c^*)$ under a known prompt $c^*$. When the victim invokes the pipeline with any prompt $c_{\text{user}}$, the model receives $\boldsymbol{z}_T^*$ instead of fresh randomness, and the output $f_\theta(\boldsymbol{z}_T^*, c_{\text{user}})$ is dictated by the attacker's chosen latent.

\textbf{Seeded-Generator Hijack.} For scenarios where $\eta > 0$ (non-deterministic sampling), additional random noise $\boldsymbol{\epsilon}_t$ is injected at each denoising step (Eq.~\ref{eq:ddim}). The seeded-generator variant addresses this by injecting a deterministic random number generator object initialized with a fixed seed $s^*$:
\begin{equation}
g = \mathrm{RNG}(\text{seed} = s^*)
\label{eq:seeded}
\end{equation}
This controls all stochastic calls throughout the denoising chain, including $\boldsymbol{z}_T$ and all intermediate noise injections $\{\boldsymbol{\epsilon}_t\}_{t=1}^{T}$. The entire generation process becomes reproducible regardless of the $\eta$ value.

\subsection{Attack Variants.}
We define two operational attack variants based on attacker knowledge:

\textbf{Attack B (Trigger-Based).} The attacker knows a target prompt $c^*$ and pre-computes $\boldsymbol{z}_T^*$ such that $f_\theta(\boldsymbol{z}_T^*, c^*) = x_0^*$ produces a desired harmful image $x_0^*$. When the victim uses prompt $c^*$, pixel-perfect reproduction is achieved (SSIM $= 1.00$). This variant requires the attacker to anticipate the victim's prompt.

\textbf{Attack A (Prompt-Agnostic).} The attacker does not know the victim's prompt. By fixing $\boldsymbol{z}_T^*$, the attacker constrains the generation to a narrow manifold region regardless of $c_{\text{user}}$. While bit-exact replication is not guaranteed (the output depends on the interaction between $\boldsymbol{z}_T^*$ and $c_{\text{user}}$), the attack significantly biases generation toward the attacker's pre-determined content, achieving elevated SSIM values ($0.222$--$0.279$ at various guidance scales, compared to random baselines of $0.066$--$0.109$) across all tested prompts.

\subsection{Inherent Undetectability.}

A critical property of DiffusionHijack is its inherent undetectability against existing defense mechanisms. Unlike conventional backdoor attacks that modify model weights or adversarial attacks that perturb inputs, our attack operates at the system supply-chain level, rendering it invisible to standard detection approaches:

\begin{itemize}
\item \textbf{No weight modification.} The model parameters remain entirely untouched. Defense methods based on weight auditing, parameter distribution analysis, or activation anomaly detection are fundamentally ineffective.

\item \textbf{No input/output perturbation.} The user prompt is unaltered, and the generated images are high-quality, semantically consistent outputs. As demonstrated in Section~\ref{sec:experiments}, the attack achieves a Safety Checker bypass rate of 98--100\%, confirming that even automated content moderation systems cannot distinguish hijacked outputs from benign ones.

\item \textbf{Attack surface below the model layer.} The attack resides in the runtime environment (runtime function replacement of the standard normal sampling function), not in the neural network itself. Every intermediate computation---latent representations, attention maps, and logits---follows the standard inference trajectory, leaving no detectable anomalous fingerprint.
\end{itemize}

This combination of properties makes DiffusionHijack fundamentally more stealthy than existing attack paradigms, as it bypasses all three layers of conventional AI security: model-level auditing, input-level filtering, and output-level content moderation.

% ============================================================
% IV. QRNG Defense
% ============================================================
\section{QRNG Defense}

\subsection{Defense Principle.}
The success of DiffusionHijack fundamentally depends on the predictability of PRNG outputs. A PRNG is a deterministic algorithm: given the same internal state, it produces identical output sequences. The attacker exploits this by either fixing the output directly (fixed-tensor hijack) or controlling the internal state via a seed (seeded-generator hijack). The defense strategy is therefore to replace the PRNG with a source of true randomness that is physically impossible to predict or reproduce.

QRNGs derive randomness from quantum mechanical processes whose outcomes are fundamentally indeterminate according to the laws of physics. No computational algorithm, regardless of its resources, can predict or reproduce the output of a properly implemented QRNG. This provides an information-theoretic security guarantee that is qualitatively stronger than any computational assumption underlying PRNGs. Unlike algorithmic defenses that can only mitigate specific attack patterns, QRNG provides information-theoretic security that is unconditionally immune to any computational adversary.

\subsection{Implementation.}
We implement the QRNG defense using a hardware QRNG integrated into the inference server. The defense pipeline operates as follows:

\textbf{Step 1: Raw Random Number Acquisition.} The QRNG hardware continuously generates uniformly distributed random numbers $u_i \sim \mathcal{U}(0, 1)$, buffered in a pre-allocated pool of $5 \times 10^7$ float64 values.

\textbf{Step 2: Inverse CDF Transform.} Uniform samples are converted to standard Gaussian samples via the inverse cumulative distribution function (CDF):
\begin{equation}
z_i = \Phi^{-1}(u_i) \sim \mathcal{N}(0, 1), \quad \forall\, i = 1, \ldots, d
\label{eq:icdf}
\end{equation}
where $\Phi^{-1}$ denotes the quantile function of the standard normal distribution.

\textbf{Step 3: Tensor Construction.} Gaussian samples are reshaped into the required latent tensor dimensions:
\begin{equation}
\boldsymbol{z}_T = \text{reshape}(\{z_i\}_{i=1}^{d}, [1, 4, H/8, W/8])
\label{eq:reshape}
\end{equation}
where $H$ and $W$ are the target image dimensions, and the factor of $8$ corresponds to the Variational Autoencoder (VAE) downsampling ratio in latent diffusion models~\cite{rombach2022ldm}.

\textbf{Step 4: Pipeline Injection.} The QRNG-generated latent is injected directly into the diffusion pipeline via the latents parameter, bypassing the internal PRNG call entirely:
\begin{equation}
x_0 = f_\theta(\boldsymbol{z}_T^{\mathrm{QRNG}},\ c_{\text{user}})
\label{eq:inject}
\end{equation}
This bypasses any compromised PRNG sampling function, as the externally supplied quantum-random latent is used directly without invoking the framework's built-in random number generation.

\subsection{Performance Overhead Analysis.}
The QRNG defense introduces negligible computational overhead relative to the diffusion inference process. The pre-allocated buffer pool of $5 \times 10^7$ float64 values occupies approximately 400~MB of system memory; initial loading of the pre-generated buffer from storage requires $< 1$~s. During inference, latent tensor construction consists of a memory read from the pre-loaded pool followed by the inverse-CDF transformation (Eq.~\ref{eq:icdf}), which completes in under 1~ms---negligible compared to the 3--5~s required for a typical 50-step DDIM denoising pass at $512 \times 512$ resolution. From a throughput perspective, a single latent tensor for $512 \times 512$ generation requires $4 \times 64 \times 64 \times 4$~bytes $\approx 65$~KB of random data, while the QRNG hardware sustains 600~Mbps (75~MB/s), comfortably supporting batch-parallel generation with over $1000\times$ headroom for single-image latent requirements. Consequently, the QRNG defense does not constitute a practical bottleneck in production inference pipelines.

\subsection{Security Analysis.}
We state the security guarantee of the QRNG defense formally.

\textbf{Theorem 1.} \textit{Under QRNG defense, the probability that an attacker's pre-computed latent $\boldsymbol{z}_T^*$ coincides with the inference latent $\boldsymbol{z}_T^{\mathrm{QRNG}}$ is zero, regardless of the attacker's computational resources.}

\textit{Proof.} Let $(\Omega, \mathcal{F}, \mu)$ be the probability space induced by the quantum measurement process, where $\Omega$ is the set of all possible measurement outcomes, $\mathcal{F}$ is the associated $\sigma$-algebra, and $\mu$ is the probability measure derived from the Born rule. Let $\boldsymbol{z}_T^{\mathrm{QRNG}} \in \mathbb{R}^{d}$ (where $d = 4 \times H/8 \times W/8$) denote the latent tensor produced by the QRNG via inverse-CDF transformation $\Phi^{-1}: [0,1] \to \mathbb{R}$ applied to uniformly distributed quantum measurement outcomes.

By the Born rule, $\mu$ is absolutely continuous with respect to the Lebesgue measure on $\mathbb{R}^d$, and each measurement outcome is fundamentally non-deterministic and statistically independent of all classical information accessible to the attacker. For any fixed point $\boldsymbol{z}_T^* \in \mathbb{R}^d$ pre-computed by the attacker, the Lebesgue measure of the singleton set $\{\boldsymbol{z}_T^*\}$ is zero, hence $\Pr[\boldsymbol{z}_T^{\mathrm{QRNG}} = \boldsymbol{z}_T^*] = \mu(\{\omega \in \Omega : \boldsymbol{z}_T^{\mathrm{QRNG}}(\omega) = \boldsymbol{z}_T^*\}) = 0$. Furthermore, causal independence of the quantum source precludes adaptive strategies: the attacker cannot condition $\boldsymbol{z}_T^*$ on future QRNG outputs, as this would violate the no-signaling principle of quantum mechanics. The guarantee holds unconditionally, without reliance on computational hardness assumptions.

\textit{Remark (Finite-Precision Bound).} In practice, floating-point representation (float64) discretizes $\mathbb{R}^d$ into at most $2^{64d}$ representable points. For a typical $512 \times 512$ latent tensor ($d = 4 \times 64 \times 64 = 16384$ elements), the collision probability is upper-bounded by $2^{-64d} = 2^{-1048576}$, a value negligible beyond any physical threshold.

\subsection{Applicability Beyond Diffusion Models.}
Although demonstrated on diffusion models in this work, the QRNG defense is architecture-agnostic and directly applicable to any generative system relying on PRNG-based sampling. This includes large language models (LLMs) that use stochastic token sampling with temperature-based decoding, variational autoencoders (VAEs) that sample from learned latent distributions via the reparameterization trick, generative adversarial networks (GANs) that draw noise vectors from prior distributions, and video diffusion models that require spatiotemporal noise tensors. In each case, replacing the PRNG with a QRNG at the sampling interface provides the same information-theoretic guarantee: no computational adversary can predict or reproduce the random inputs to the generative process, thereby neutralizing any supply-chain attack that targets the randomness infrastructure.

% ============================================================
% V. Experiments
% ============================================================
\section{Experiments}
\label{sec:experiments}

\subsection{Experimental Setup.}
All experiments were conducted on a workstation equipped with an NVIDIA RTX 3090 GPU (24~GB VRAM) running Ubuntu 20.04. The software environment consisted of Python 3.10, PyTorch 2.1.0 with CUDA 12.1, and the Diffusers library v0.25.0 for model loading and inference pipeline management. The QRNG hardware was a Silicon Extreme QRNG600 PCIe card (Hefei GZ-iChip Technology Co., Ltd.) providing 600~Mbps of quantum random data, with a pre-allocated buffer of $5 \times 10^7$ float64 Gaussian samples (pre-generated offline from the QRNG hardware at 600~Mbps and loaded from storage in $< 1$~s). Single-image generation takes approximately 3--5~s for 50-step DDIM sampling at $512 \times 512$ resolution.

\textbf{Models.} We evaluated three diffusion models: Stable Diffusion v1.4 (SD~1.4), Stable Diffusion v1.5 (SD~1.5), and Stable Diffusion XL (SDXL)~\cite{podell2023sdxl}. All models were loaded with official pre-trained weights from Hugging Face without modification.

\textbf{Metrics.} Image similarity was quantified using Structural Similarity Index Measure (SSIM), which evaluates structural similarity between two images on a scale of $[0, 1]$, where SSIM $= 1.0$ indicates pixel-identical content. We compute SSIM using a standard $11 \times 11$ Gaussian window and average over RGB channels following the default PyTorch implementation.

\textbf{Configuration.} Unless otherwise stated, all experiments used $T = 50$ denoising steps, classifier-free guidance (CFG)~\cite{ho2022cfg} scale $w = 7.5$. A larger $w$ enforces stronger alignment with the text prompt at the cost of reduced latent diversity, while a smaller $w$ grants more freedom to the latent noise. Image resolution was $512 \times 512$ (SD~1.4/1.5) or $1024 \times 1024$ (SDXL). Each condition was evaluated across $K = 10$ diverse text prompts with $n_{\text{rep}} = 10$ independent repetitions per prompt, yielding $N = 100$ total trials per experimental condition.

\subsection{Main Results.}
Table~\ref{tab:main} presents the primary attack and defense evaluation results across all three models.

\begin{table}[!t]
\centering
\caption{Comprehensive attack and defense evaluation. SSIM reported as mean $\pm$ std over $N = 100$ trials ($K = 10$ prompts $\times$ $n_{\text{rep}} = 10$ repetitions) per condition. All experiments use $T = 50$ steps, $w = 7.5$.}
\label{tab:main}
\label{tab:eta}
\renewcommand{\arraystretch}{1.2}
\setlength{\tabcolsep}{3pt}
\footnotesize
\begin{tabular}{@{}ll ccc@{}}
\toprule
\textbf{Model} & \textbf{$\eta$} & \textbf{Attack} & \textbf{Baseline} & \textbf{QRNG Defense} \\
\midrule
SD 1.4 & 0.0 & $1.000 \pm 0.000$ & $0.186 \pm 0.086$ & $0.165 \pm 0.094$ \\
SD 1.5 & 0.0 & $1.000 \pm 0.000$ & $0.179 \pm 0.072$ & $0.183 \pm 0.133$ \\
SD 1.5 & 0.2 & $1.000 \pm 0.000$ & $0.182 \pm 0.075$ & $0.180 \pm 0.128$ \\
SD 1.5 & 0.5 & $1.000 \pm 0.000$ & $0.185 \pm 0.081$ & $0.177 \pm 0.121$ \\
SDXL   & 0.0 & $1.000 \pm 0.000$ & $0.400 \pm 0.103$ & $0.402 \pm 0.130$ \\
\bottomrule
\end{tabular}
\end{table}

The attack achieves perfect reproduction (SSIM $= 1.00$) with zero variance across all trials, confirming deterministic output under DDIM. The QRNG defense reduces SSIM to levels statistically indistinguishable from the unmodified baseline (two-sample $t$-test, $p > 0.05$ for all model pairs), demonstrating complete attack neutralization without degrading generation diversity.

\begin{figure*}[!t]
\centering
\includegraphics[width=\textwidth]{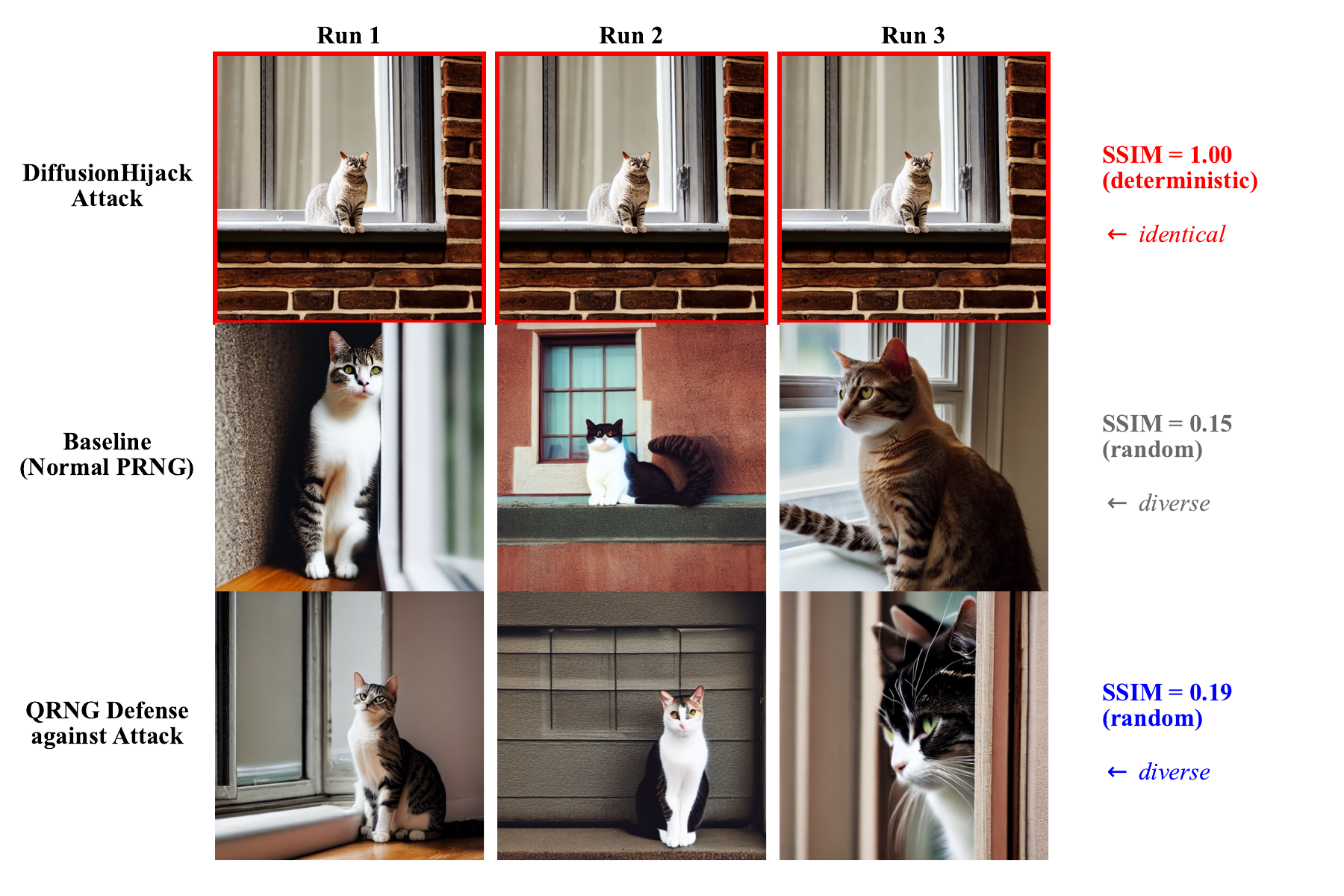}
\caption{Visual comparison of DiffusionHijack attack, baseline, and QRNG defense outputs on SD~1.5. Three independent runs are shown for the same prompt (``a photo of a cat sitting on a windowsill''). Top row (red border): DiffusionHijack forces pixel-identical outputs (SSIM $= 1.00$). Middle row: unmodified PRNG produces diverse images (SSIM $= 0.15$). Bottom row: QRNG-sourced latents restore generation diversity (SSIM $= 0.19$), confirming complete attack neutralization.}
\label{fig:visual_comparison}
\end{figure*}

\begin{figure}[!t]
\centering
\includegraphics[width=\columnwidth]{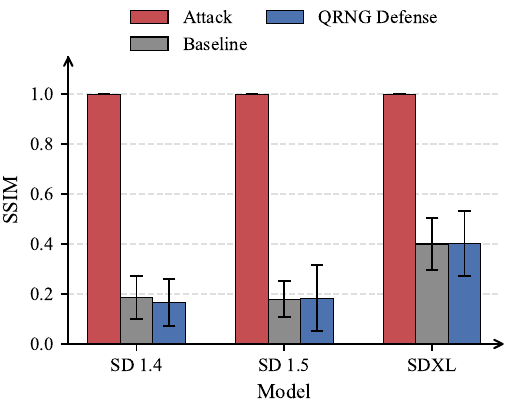}
\caption{Cross-model SSIM comparison under attack, baseline, and QRNG defense conditions. Results are shown for SD~v1.4, v1.5, and SDXL (DDIM, $\eta = 0$, $T = 50$, $w = 7.5$, $N = 100$ trials per condition). DiffusionHijack achieves SSIM $= 1.000 \pm 0.000$ across all models. Baseline SSIM ranges from $0.179 \pm 0.072$ (SD~1.5) to $0.400 \pm 0.103$ (SDXL). QRNG defense yields SSIM statistically indistinguishable from baseline ($p > 0.05$, two-sample $t$-test). Legend---red: DiffusionHijack; gray: baseline; blue: QRNG defense. Error bars indicate $\pm 1\sigma$.}
\label{fig:cross_model_ssim}
\end{figure}

The elevated baseline SSIM for SDXL ($0.400 \pm 0.103$) compared to SD~1.4/1.5 ($\sim$0.18) reflects SDXL's higher structural consistency and is unrelated to any attack effect. Fig.~\ref{fig:visual_comparison} provides a visual comparison: under the same prompt (``a photo of a cat sitting on a windowsill''), three independent runs of the attacked pipeline produce pixel-identical outputs (SSIM $= 1.00$), whereas both the normal baseline and QRNG-defended pipelines yield diverse, prompt-appropriate images (SSIM $\approx 0.15$--$0.19$).

\subsection{Robustness Under Non-Deterministic Sampling.}
Table~\ref{tab:main} also reports the seeded-generator hijack under non-deterministic sampling conditions with varying noise injection levels ($\eta \in \{0.2, 0.5\}$). The seeded-generator hijack maintains perfect reproduction (SSIM $= 1.00$) even when $\eta > 0$, because the fixed generator seed controls all stochastic calls in the denoising chain. The attack is therefore not limited to deterministic samplers.

\subsection{Safety Checker Bypass.}
Table~\ref{tab:safety} evaluates the attack's ability to bypass CLIP-based safety mechanisms deployed in Stable Diffusion pipelines.

\begin{table}[!t]
\centering
\caption{Safety checker bypass on SD~1.5. Bypass rate: fraction evading the CLIP classifier. CLIP shift: cosine distance between user prompt embedding and generated image embedding ($N = 50$).}
\label{tab:safety}
\renewcommand{\arraystretch}{1.2}
\begin{tabular}{lcc}
\toprule
\textbf{Attack Type} & \textbf{Bypass Rate (\%)} & \textbf{CLIP Shift} \\
\midrule
Fixed-tensor (NSFW target) & 100.0 & $2.6 \pm 0.3$ \\
Seeded-generator (NSFW target) & 98.0 & $2.2 \pm 0.4$ \\
\bottomrule
\end{tabular}
\end{table}

The CLIP-based safety checker uses pre-trained CLIP image-text embeddings to detect inappropriate content by comparing generated image features against a set of known unsafe concept embeddings; images whose similarity exceeds a threshold are blocked. The attack achieves 98--100\% bypass rates. The hijacked latent forces a pre-determined output irrespective of the prompt, and the attacker can select latents not flagged by the classifier's threshold. CLIP shift values of $2.2$--$2.6$ confirm substantial deviation from the user's intended content.

\subsection{Prompt-Agnostic Attack.}
Table~\ref{tab:agnostic} evaluates Attack~A, where the attacker fixes $\boldsymbol{z}_T^*$ without knowledge of the victim's prompt.

\begin{table}[!t]
\centering
\caption{Prompt-agnostic attack (Attack~A) on SD~1.5 across CFG scales $w$ ($T = 50$, $\eta = 0$, $N = 100$ prompt pairs per $w$). Reduction is defined as $(1 - \text{SSIM}_{\text{QRNG}} / \text{SSIM}_{\text{Attack}}) \times 100\%$.}
\label{tab:agnostic}
\renewcommand{\arraystretch}{1.2}
\begin{tabular}{lccc}
\toprule
\textbf{Guidance $w$} & \textbf{Attack SSIM} & \textbf{QRNG Defense} & \textbf{Reduction} \\
\midrule
1.0 & $0.279$ & $0.066$ & 76\% \\
3.0 & $0.256$ & $0.091$ & 64\% \\
7.5 & $0.222$ & $0.109$ & 51\% \\
\bottomrule
\end{tabular}
\end{table}

At low guidance scales ($w = 1.0$), the model relies more heavily on the latent structure, resulting in higher attack effectiveness ($\text{SSIM} = 0.279$). As guidance scale increases, the text conditioning exerts stronger influence on the output, reducing the attacker's control. Nevertheless, even at $w = 7.5$, the attack SSIM ($0.222$) remains substantially above the QRNG-defended baseline ($0.109$), representing a $2.0\times$ elevation. The QRNG defense consistently reduces attack effectiveness by 51--76\% across all guidance scales, confirming its robustness as a countermeasure.

\begin{figure}[!t]
\centering
\includegraphics[width=\columnwidth]{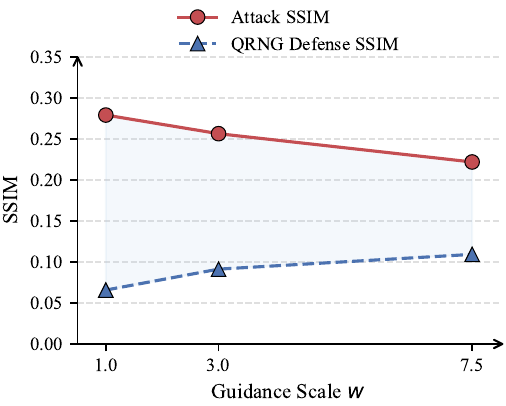}
\caption{Effect of CFG scale $w$ on prompt-agnostic attack effectiveness (SD~1.5). Red solid line with circle markers: Attack SSIM; blue dashed line with triangle markers: QRNG Defense SSIM. Shaded region indicates defense reduction magnitude. Conditions: $T = 50$, $\eta = 0$, $N = 100$ prompt pairs per scale.}
\label{fig:guidance_scale}
\end{figure}

% ============================================================
% VI. Discussion
% ============================================================
\section{Discussion}

\subsection{Real-World Feasibility.}
The supply-chain attack vector demonstrated in this work is realistic and practically achievable. The Python Package Index (PyPI) has documented numerous instances of malicious packages employing typo-squatting and dependency confusion~\cite{williams2024supplychain}. In the machine learning ecosystem, where practitioners frequently install packages with minimal verification, a malicious fork of the diffusion inference library or a poisoned dependency of the deep learning framework could inject PRNG manipulation code transparently. Docker images hosted on public registries present a similar vector, as users typically do not audit the runtime environment. The attack requires only a few lines of code to implement (runtime function replacement of the normal sampling function or injecting a seeded generator), making it readily concealable within a large codebase.

\subsection{Limitations.}
Several limitations of the current work warrant discussion. First, the attack requires supply-chain-level access to the victim's inference environment, which, while realistic, is more difficult to achieve than purely remote attacks. Second, the QRNG defense requires dedicated hardware, which incurs additional cost and may not be feasible for all deployment scenarios. Third, our evaluation is limited to the Stable Diffusion family; while the attack principle generalizes to any PRNG-dependent generative model, empirical validation on other architectures (e.g., DALL$\cdot$E~3, Imagen) remains future work. Fourth, the prompt-agnostic attack achieves moderate rather than perfect similarity, suggesting that stronger conditioning mechanisms may partially mitigate this variant without dedicated defense.

\subsection{Lightweight Alternatives and Engineering Trade-offs.}
While QRNG provides the strongest theoretical guarantee, practical deployments may consider cost-effective alternatives that offer intermediate security levels. System-level hardware entropy sources, such as the operating system kernel entropy pool (backed by hardware interrupt timing and device driver noise) and CPU-integrated random number instruction sets (e.g., Intel RDRAND/RDSEED), provide non-deterministic randomness derived from physical phenomena at negligible cost. These sources are immune to the fixed-seed and fixed-tensor hijack variants, as their outputs cannot be predicted or reproduced by an external adversary. Distributed true random number aggregation services, which combine entropy from multiple geographically dispersed hardware sources, offer another scalable option with enhanced tamper resistance. However, these alternatives carry important caveats: system-level entropy pools may be vulnerable to state recovery attacks under certain threat models~\cite{dodis2013entropy}, and CPU-integrated generators rely on manufacturer trust assumptions that dedicated QRNGs avoid. We position these as pragmatic engineering trade-offs: they provide security substantially above pure algorithmic PRNGs but below the unconditional guarantee of dedicated QRNGs, and are appropriate for deployments where the full quantum hardware investment is not justified by the threat level.

\subsection{Broader Implications.}

\begin{table}[t]
\centering
\caption{Detectability comparison across attack paradigms. \ding{51}: yes; \ding{55}: no.}
\label{tab:detectability}
\renewcommand{\arraystretch}{1.3}
\small
\begin{tabular}{l c c c c}
\toprule
\textbf{Attack Type} & \textbf{\makecell{Weight\\Mod.}} & \textbf{\makecell{Input\\Mod.}} & \textbf{\makecell{Weight\\Audit}} & \textbf{\makecell{I/O\\Filter}} \\
\midrule
DiffusionHijack (Ours) & \ding{55} & \ding{55} & \ding{55} & \ding{55} \\
Training-time poisoning~\cite{jang2025silent} & \ding{51} & \ding{55} & \ding{51} & \ding{55} \\
Fine-tuning backdoor~\cite{zhai2023baadt2i}    & \ding{51} & \ding{55} & \ding{51} & \ding{55} \\
Prompt injection~\cite{zhang2025reason2attack}        & \ding{55} & \ding{51} & \ding{55} & \ding{51} \\
Adversarial examples~\cite{dai2024advdiff}    & \ding{55} & \ding{51} & \ding{55} & \ding{51} \\
\bottomrule
\end{tabular}
\end{table}

Our findings have implications beyond diffusion models. Any generative model that relies on PRNGs for stochastic sampling---including large language models (LLMs) using temperature-based decoding, variational autoencoders, and generative adversarial networks---may be vulnerable to analogous PRNG manipulation attacks. In particular, the PRNG supply chain of LLM inference servers represents a similarly unaudited attack surface where seed control could bias token sampling toward attacker-chosen outputs, suggesting a unified threat paradigm across generative AI modalities. We advocate for:
\begin{enumerate}
  \item Systematic security audits of AI software supply chains, with particular attention to randomness generation components.
  \item Integration of hardware-based true random number generation into critical AI infrastructure.
  \item Development of runtime integrity verification mechanisms for PRNG behavior.
\end{enumerate}

% ============================================================
% VII. Conclusion
% ============================================================
\section{Conclusion}
This paper presented DiffusionHijack, a supply-chain backdoor attack that hijacks the PRNG in diffusion model inference to achieve deterministic output control. The attack achieves deterministic output duplication (SSIM $= 1.00$) across three Stable Diffusion variants, persists under stochastic sampling, bypasses CLIP safety checkers (98--100\%), and enables prompt-agnostic forced generation. As a countermeasure, we replace the PRNG with a hardware QRNG that provides information-theoretic unpredictability, completely neutralizing the attack.

Our work reveals a critical yet overlooked vulnerability in the generative AI supply chain and demonstrates that hardware-based quantum randomness offers an effective, unconditional defense. We call on the community to prioritize supply-chain security audits for AI systems, with particular attention to the randomness infrastructure underpinning all stochastic models. Future work will extend the evaluation to additional generative architectures (e.g., DALL$\cdot$E~3, video diffusion) and investigate lightweight QRNG integration strategies suitable for edge deployment.

\section*{Acknowledgment}

This work was supported by the National Natural Science Foundation of China (72573124).

% ============================================================
% References (inlined from .bbl for arXiv submission)

\end{document}